\documentstyle[prl,aps,multicol,epsf]{revtex}
\newcommand{\notAPS}[1]{#1}

\newcommand{\RM}{\notAPS{\rm}}
\newcommand{\journ}[5]{{\notAPS{\sl}#1} {\bf #2}, #4 (#3)}

\newcommand{\spin}{1} 
\newcommand{\cond}{2} 
\newcommand{\figwidth}{8.1cm}
\newcommand{\ket}[1]{|#1\rangle}
\newcommand{\vac}{\ket{0}}
\newcommand{\ii}{{\RM i}}
\newcommand{\eee}{{\RM e}}
\newcommand{\tots}{{\cal S}} 
\newcommand{\tJ}{$t$-$J$}
\newcommand{\fig}{Fig.}
\newcommand{\vF}{v_{\RM F}}
\newcommand{\Tc}{T_{\RM c}}
\newcommand{\JK}{J_{\RM K}}
\newcommand{\JJ}[2]{{\vec{\rho}}_{{\RM #1}}^{{\:#2}}}
\newcommand{\JH}{J_{\RM H}}
\newcommand{\hlf}{{\textstyle\frac{1}{2}}}
\newcommand{\eq}[1]{Eq.\ (\ref{#1})}
\newcommand{\bea}{\begin{eqnarray}}
\newcommand{\eea}{\end{eqnarray}}
\newcommand{\be}{\begin{equation}}
\newcommand{\ee}{\end{equation}}
\newcommand{\kF}{k_{\RM F}}
\newcommand{\Tr}{{\rm tr}}
\newcommand{\hc}{{\rm h.c.}}
\newcommand{\dg}{^{\dagger}}
\newcommand{\gdg}{^{\ }}     
\newcommand{\up}{\uparrow}
\newcommand{\dn}{\downarrow}
\newcommand{\nn}[1]{\langle #1 \rangle}
\newcommand{\nnij}{\nn{ij}}
\newcommand{\etal}{{\it et al.}}

\begin{document} 
\draft 
\title{Spin Gap in a Doped Kondo Chain}
\author{
Arnold E. Sikkema$^{1,2}$,
Ian Affleck$^{2,3}$,
Steven R. White$^4$
}
\address{
$^1$Institute for Fundamental Theory,
Department of Physics,
University of Florida,
Gainesville, FL 32611-8440\\
$^2$Department of Physics \& Astronomy and
$^3$Canadian Institute for Advanced Research,\\
University of British Columbia,
6224 Agricultural Road,
Vancouver, BC,
Canada~~V6T~1Z1\\
$^4$Department of Physics \& Astronomy,
University of California,
Irvine, CA 92697}

\date{14 February 1997; revised 26 March 1997}
\maketitle

\begin{abstract}
We show that the Kondo chain away from half-filling has a spin gap
upon the introduction of an additional direct
Heisenberg coupling between localized spins.  This is understood in
the weak-Kondo-coupling limit of the Heisenberg-Kondo lattice model by
bosonization and in the strong-coupling limit by a mapping to a
modified \tJ\ model.  Only for certain ranges of filling and
Heisenberg coupling does the spin gap phase extend from weak to
strong coupling.
\end{abstract} 

\begin{multicols}{2}
\narrowtext

The Kondo lattice model (KLM)
[\eq{HKLM-Hamiltonian} with $\JH=0$]
is
often
considered as a model for
heavy fermion materials. The one dimensional (1D) version
has recently
been argued to be very relevant to a theoretical approach to
high-$\Tc$ superconductivity based on fluctuating hole-rich stripes
\cite{EKZ}.  It has been rather clearly established that, at
half-filling, the 1D KLM has a gap to both spin and charge excitations
\cite{JullienPfeuty_Tsunetsugu_YuWhite}.  On the other hand, its behavior away
from half-filling remains controversial.  Of particular
interest is the possibility of a phase with a spin gap but no charge
gap.  In this case, the charge density wave and superconducting
susceptibilities are significantly enhanced, and for effectively
attractive interactions the system is expected to become
superconducting upon the introduction of any inter-chain hopping.   

A large Kondo coupling forces all conduction electrons to form on-site
singlets with localized spins.  The unpaired localized spins then
effectively hop on the background of singlets via a reduced hopping
term $t/2$.  Additional $t^2/\JK$ interactions then lead to incomplete
ferromagnetism \cite{Sigrist92}. At weaker Kondo coupling a Luttinger
liquid (LL) phase has been proposed and supported, to some extent, by
density matrix renormalization group (DMRG) work
\cite{MoukouriCaron,Shibata}.

Part of the difficulty in understanding this model is that it cannot
be bosonized in a very direct manner.  Two approaches have been
proposed.  In one of them \cite{ZKE} a special, ``Toulouse'', limit
with a highly anisotropic Kondo interaction is bosonized, resulting in
a spatially modulated sine-Gordon model and leading to the prediction
of a spin gap (SG) phase with a spin gap proportional to the ${2\over3}$
power of the perpendicular part of the Kondo coupling.

An alternative approach can be applied to a  more general
Heisenberg-Kondo lattice model (HKLM) which has a direct Heisenberg
coupling $\JH$ between the localized spins as well as the usual Kondo
interaction:  
\be
\label{HKLM-Hamiltonian}
H=
-t \sum_{\nnij\sigma} c\dg_{i\sigma}c\gdg_{j\sigma}+\hc
+\JK \sum_i \vec{S}_i\cdot \vec{s}_i
+\JH \sum_{\nnij} \vec{S}_i\cdot\vec{S}_j.
\ee
Here $c\dg_{i\sigma}$  is a creation operator for an electron of spin
$\sigma$ at site $i$,
$\vec s_i=\sum_{\alpha\beta}c\dg_{i\alpha}\hlf\vec\sigma_{\alpha\beta}c\gdg_{i\beta}$
is the conduction
electron spin operator, $\vec{S}_i$ is the localized
spin-$\hlf$
operator, and $\nnij$ denotes nearest neighbors.  This model can be
bosonized \cite{FujimotoKawakami,WhiteAffleck} in the limit $\JK\ll t$,
$\JH$ in which only the low energy degrees of freedom of the
Heisenberg chain and the conducting free electron chain are kept.
This approach predicts \cite{WhiteAffleck} a SG phase.
Another interesting limit of the generalized
model \cite{WhiteAffleck,SikkemaThesis}
is $\JK\gg t$ and $\JK\gg\JH\gg t^2/\JK$ in which it becomes equivalent to a
\tJ\ model with an effective electron density of
$n_{\rm eff}=|1-n|$, where $n$ is the conduction electron density.
Building on the rather well understood phase diagram of the \tJ\ model
we learn that LL and phase-separated (PS) regimes exist as well as a
SG phase for some range of $n$ near half-filling.  Thus we see that
there must be at least four phases in the HKLM.  

We note that the usual KLM contains an effective RKKY interaction
between the localized spins, but this would normally be of longer range
and weaker than our $\JH$.  Such a short-range exchange interaction is
believed to be present, although very small, in certain
quasi-one-dimensional organic conductors containing Cu$^{2+}$ ions,
 Cu(tatbp)I and Cu(pc)I
\cite{Ogawa_Quirion}.  

The purpose of the present work is to check the analytical predictions
based on bosonization and the large $\JK$ limit using the DMRG
numerical method.  In particular,
we establish the existence of the SG phase over a
significant region of parameters.  We first briefly review the large
$\JK$ limit and the bosonization approach of Ref. \cite{WhiteAffleck},
clarifying some points about the large $\JK$ limit. 

When $\JK\gg t$ and $\JK\gg\JH\gg t^2/\JK$, 
the HKLM reduces
to the \tJ\
model \cite{WhiteAffleck,SikkemaThesis} except that the Coulomb term
$n_i n_j$ usually included in the \tJ\ model is absent:
\be
H = - t \sum_{\nnij\sigma} c\dg_{i\sigma}c\gdg_{j\sigma} +\hc
          + J\sum_{\nnij}\vec{s}_i\cdot\vec{s}_j,
\label{ModifiedtJModel}
\ee
with the re-definitions $t\to t/2$ and $n\to|1-n|$,
and $\JH$ playing the role of $J$.
The no-double-occupancy
restriction in the \tJ\ model is automatically
met by the HKLM in the limit of large $\JK$ since the
effectively-hopping spins have the same restriction.  Unlike in the
\tJ\ model, the Heisenberg coupling exists not only between the
unpaired spins; however, the connection with the \tJ\ model is seen to
be complete upon noting that the Heisenberg coupling is ineffectual
for localized spins which are already strongly coupled with the conduction
electrons in an on-site singlet pair.

\begin{figure}
\notAPS{\epsfxsize=\figwidth\centerline{\epsffile{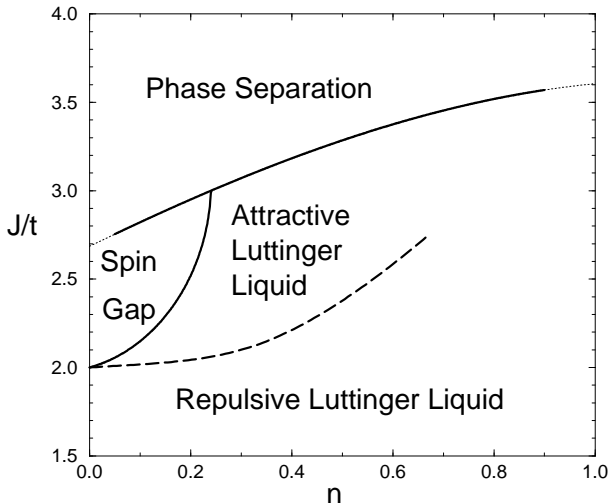}}}
\caption{A numerically-computed phase diagram of the one-dimensional
\tJ\ model (taken from Ref. \protect\cite{HellbergMele}).}
\label{fig:HellbergPD}
\end{figure}

Numerical work on the one-dimensional \tJ\ model \cite{HellbergMele}
show a spin gap phase for low electron density and $2<J/t<3$
(\fig\ \ref{fig:HellbergPD}).
This SG phase adjusts somewhat upon dropping its Coulomb term
\cite{SikkemaThesis}, as can be seen by considering the case of two
electrons in the \tJ\ model following Ref. \cite{ChenLee}.  Using
the ansatz ground state 
\be
\ket{\Psi} = \sum_{n=1}^{\infty} a^n b\dg_n \vac,
\ee
where $a$ is a constant to be determined and
$b\dg_n\equiv\sum_i(c_{i\up}\dg c_{i+n,\dn}\dg-c_{i\dn}\dg c_{i+n,\up}\dg)$
is a singlet state creation operator giving two
electrons $n$ lattice sites apart, we find that the boundary between
the LL and SG phases in the limit of zero filling
in the \tJ\ model is shifted upon dropping its Coulomb term from
$J=2t$ to $J=8t/3$, and a rough overestimate [see \fig\
\ref{fig:HellbergPD}] of that between the SG and PS
phases is shifted from $J\doteq3.2t$ to $J\doteq6.3t$
\cite{SikkemaThesis}.  Using this as a guide, we can choose values of
$\JH$ and filling which allow us to find a spin gap in the
strong-Kondo-coupling limit of the HKLM.

The phase diagram of the HKLM is three dimensional, with parameters
$\JK/t$, $\JH/t$, and the filling $n/2$.  Since filling is the most
difficult parameter to vary in our numerical DMRG study, we choose a
particular value $n={7\over8}$ to ensure that in the large $\JK$ limit we
can pass through the SG region known from the \tJ\ model by
varying $\JH$.  At large $\JK$ we expect, as we increase $\JH$ from
zero, to find first a small ferromagnetic (FM) region until the Heisenberg
term becomes comparable to the ${\cal O}(t^2/\JK)$ fluctuations
responsible for ferromagnetism, then a LL region,
followed by a SG region, and finally a PS region.

Thinking of the HKLM as a spin-$\hlf$ chain coupled to a conduction
electron chain by a weak interchain (Kondo) coupling, we proceed via
non-Abelian bosonization, writing \cite{FujimotoKawakami,WhiteAffleck} 
\be
\vec{S}_j \approx \JJ{L}{\spin} + \JJ{R}{\spin} + 
\alpha_{\spin}
(-1)^j \Tr(\vec\sigma g_{\spin})
\ee
for the spin chain, and
\be
\vec{s}_j \approx \JJ{L}{\cond} + \JJ{R}{\cond} + 
\alpha_{\cond}
[\eee^{2\ii\kF x}
\Tr(\vec\sigma g_{\cond})
\eee^{\ii\sqrt{2\pi}\phi_{\RM c}}+\hc]
\ee
for the conduction electrons.
($\alpha_{\spin,\cond}$ are constants, and
we employ the convention of using the variable $x$ for the continuum
representation of lattice position $j$.)
The SU(2) currents $\JJ{L,R}{\spin,\cond}(x)$ have scaling dimension 1,
while the SU(2) matrix fields $g_{\spin,\cond}(x)$ and the charge boson
operator $\eee^{\ii\sqrt{2\pi}\phi_{\RM c}(x)}$ have dimension $\hlf$.
At half-filling the relevant (scaling dimension ${3\over2}$) product of
alternating terms produces a gap, $\Delta_{\RM s}\propto \JK^2$. Away from
half-filling, below the length scale $1/(\pi -2k_F)$, only the
marginal interaction
\be
H_{\rm int} = \JK
(\JJ{L}{\spin}+\JJ{R}{\spin})
\cdot
(\JJ{L}{\cond}+\JJ{R}{\cond}),
\ee
survives, similar to the case of the spin-$\hlf$ spin ladder with zig-zag
rungs \cite{WhiteAffleck}.   Under renormalization group flow, this
interaction flows to zero coupling for
FM $\JK<0$, giving no gap, and to strong coupling for
antiferromagnetic $\JK>0$, giving a
spin gap
\be
\Delta_{\RM s} \propto \eee^{-c v/\JK},
\label{smallJKgap}
\ee
where $c$ is a positive constant, $v = v_{\spin} + v_{\cond}$
\cite{FujimotoKawakami},
and the two chains'
different
spin-wave velocities are
$v_{\spin} = \pi\JH/2$ and
$v_{\cond} = 2t\sin(\pi n/2)$.

All of the DMRG calculations for the HKLM reported here were done
using the finite system method \cite{WhiteDMRG}.
Spin gaps are obtained from reliable extrapolations to
infinite chain length from several different lengths
commensurate with the filling $n={7\over8}$ admitting total spin
projection
$\tots^z=0$, namely
$L=16$, 32, 48, and 64.  The calculations were done on an IBM
RS/6000 workstation equipped with 64MB RAM; typical processing times
are on the order of 20 hours of CPU time per ground state calculation
for a chain of length 64 keeping 180 states in two finite system
method sweeps, although such accuracy was not required for all points
calculated.  As a general rule, the discarded weights (sums of the
density matrix eigenvalues neglected in the basis truncation at each
step) in this case are usually on the order of $10^{-8}$ for large
Kondo coupling $\JK$ and up to $10^{-4}$ for small $\JK$ or near phase
boundaries.

\begin{figure}
\notAPS{\epsfxsize=\figwidth\centerline{\epsffile{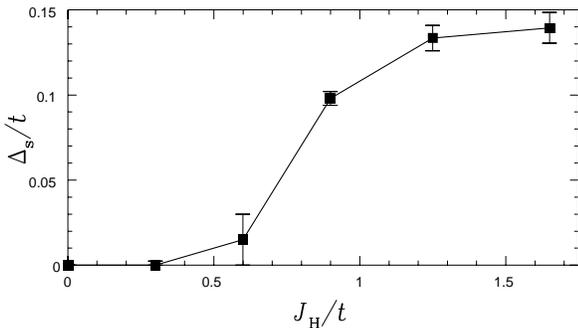}}}
\caption{The spin gap in the HKLM along $\JK=2t$ as a function of
$\JH$ shows a transition near $\JH=0.6t$.  Error bars are approximate
measures of how much we could reasonably expect the values to change
by keeping all states.}
\label{fig:JK2}
\end{figure}

Our DMRG calculations verify that there are points in the phase
diagram which have a spin gap, and points which do not.  Where these
are close enough together, we can estimate the location of the phase
boundary.  Based on our results, we find this boundary along the line
$\JK=10t$ lies between $\JH=1.25t$ and $1.65t$, and along the line
$\JH=1.25t$ it lies between $\JK=5t$ and $\JK=10t$ (close to
$\JK=7.5t$).  As shown in \fig\ \ref{fig:JK2}, the phase boundary along
the line $\JK=2t$ is near $\JH=0.6t$.

\begin{figure} 
\notAPS{\epsfxsize=\figwidth\centerline{\epsffile{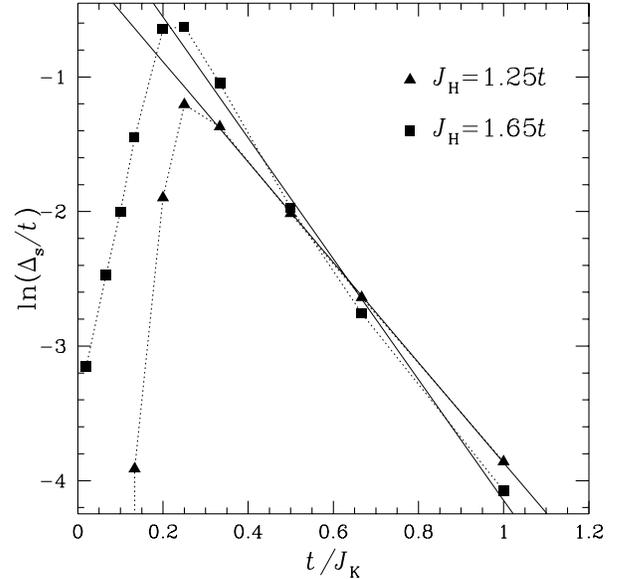}}} 
\caption{Testing the small-$\JK$ bosonization
prediction for two values of $\JH$ by plotting $\ln(\Delta_{\RM s}/t)$
versus $t/\JK$.}
\label{fig:smallJK}
\end{figure}

A test of the bosonization prediction of \eq{smallJKgap} is depicted
in \fig\ \ref{fig:smallJK}.  
For large $\JK$ the exponential prediction is not applicable
and in fact the gap goes through a maximum in the approach to the
large $\JK$ limit which maps onto a modified \tJ\ model.  The
resulting approximate linearity in the range $t\le\JK\le3t$ is good
verification of the bosonization prediction. The slope
is to be compared with $-cv$;
the ratio of the
slopes for the two values of $\JH$ is predicted to be 0.86, while it
is numerically determined to be $0.83\pm10$\%.  The value of $c$
determined from the DMRG calculations is $1.1\pm10$\%.

\begin{figure}
\notAPS{\epsfxsize=\figwidth\centerline{\epsffile{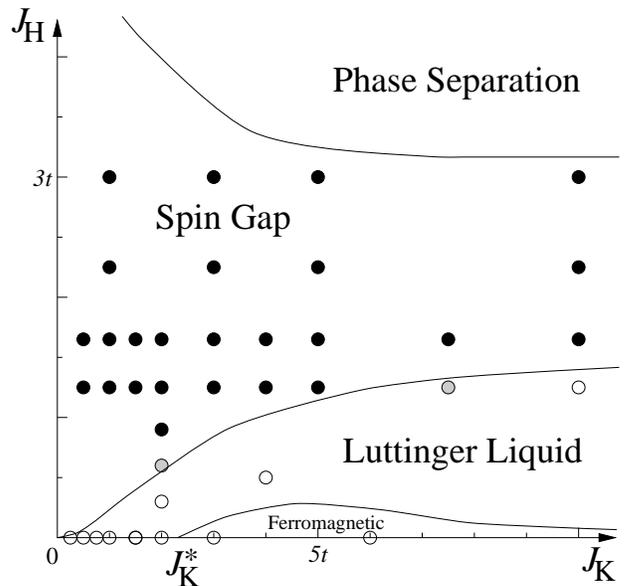}}}
\caption{A possible phase diagram of the one-dimensional HKLM at
$n={7\over8}$.
Filled circles are points at which a finite spin gap was
measured by DMRG, open circles are gapless points (or have spin gaps less
than 0.5\% of $t$), and shaded circles
are too close to a phase boundary to determine the gap.  For 
fillings further from $n=1$, the spin gap phase at large $\JK$ shrinks and
eventually closes off, while remaining at smaller $\JK$.}
\label{fig:Phases}
\end{figure}

We are thus led to conjecture a phase diagram for the HKLM, near
half-filling, like the one shown in \fig\ \ref{fig:Phases}.  We can
regard the large $\JK$ region of this diagram as being rather firmly
established by prior work on the \tJ\ model. We have also assumed the
correctness of the bosonization prediction of a spin gap at small
$\JK$ (for all non-zero $\JH$), encouraged by the above DMRG
verification.  This DMRG work also suggests that the SG
phase at large $\JK$ is connected to the one at small $\JK$.  The
boundary of the FM phase at a finite $\JK^*$ has also been
established by previous work.  We expect this FM phase to
persist, for large enough $\JK$, up to some finite $\JH$ where the
predominant interaction between unpaired localized spins becomes
antiferromagnetic.  Possibly the most interesting phase boundary is
the one between LL and SG phases.  This
could extend all the way to $\JK=0$ as drawn in \fig\
\ref{fig:Phases}, in which case the SG phase would not
occur in the pure Kondo lattice model.  Alternatively it could
intercept the $\JK$ axis at a finite value, $\JK^{*2}<\JK^*$.
Finally, it might intercept the boundary of the FM region
at finite $\JH$ in which case the pure KLM would make a transition
directly from the FM phase into the SG phase
without passing through the LL phase.  In the latter two
scenarios, the spin gap in the pure KLM should turn on below $\JK^*$
or $\JK^{*2}$, pass through a maximum, and then vanish at $\JK\to 0$
where the model becomes non-interacting.  If this maximum spin gap is
not too small (greater than a few percent of $t$) it should be
observable by DMRG.
An important parameter for the spin excitations in either LL or SG
phase is the spin-wave velocity, $v_{\RM s}$.  In the pure KLM, we expect
this to go to zero at the boundary of the FM phase, pass through a
maximum as $\JK$ decreases, and then vanish as $\JK\to 0$.  In the
second scenario, it should remain finite at $\JK^{*2}$, where the
spin gap turns on.  We expect that the spin gap will not exceed
$v_{\RM s}$ in order of magnitude.  In regions where $v_{\RM s}$ is very small
(which may include the entire $\JK$ axis  below $\JK^*$)
the LL (or SG) behavior only appears at very low energies (and
long lengths) and hence is difficult to study numerically. 
 A non-zero $\JH$ enhances $v_{\RM s}$, making it easier to determine
the phase diagram.    So far, we have 
not found
a spin gap in the pure KLM, although we have not exhaustively
studied all values of $\JK$ and $n$.  For $n={7\over8}$, we see evidence of
a FM/LL phase boundary between $\JK/t=1$ and 2; no spin gap is seen
down to $\JK/t={1\over4}$.
This is consistent 
with what was found earlier by DMRG in a somewhat different parameter
regime \cite{Shibata}.
For lower densities, the FM/LL boundary
moves closer to $\JK=0$, since in the limit of $n\to0$ the ground
state of the KLM has been rigorously shown to be FM
\cite{Sigrist91}, closing the window of opportunity for a SG phase in
the KLM.  At densities lower than about $n={3\over4}$, we know from work on
the \tJ\ model that the SG phase doesn't exist at large $\JK$.
Nevertheless, assuming the veracity of the bosonization approach, we
might expect the SG phase to exist at non-zero $\JH$ and
smaller $\JK$.

The above bosonization approach implies that the SG phase
is a charge-only LL.  This conclusion is supported by
the fact that we find numerically that the charge-transfer gap
$\Delta_{\RM ct}(n,L)\equiv E_L(nL+2)+E_L(nL-2)-2E_L(nL)$, where
$E_L(Q)$ is the ground state energy of the length-$L$ chain in the
sector of charge $Q$ and $\tots^z=0$, is zero in
the thermodynamic limit in this phase.

A conformal field theoretic finite-size scaling analysis predicts that
in a charge-only LL, the coefficient of $Q^2$ in the
energy of the ground state in the sector of charge $Q$ should be
${\pi v_{\RM c}\over2L}\cdot{1\over4\pi R^2}+\cdots$ where $\cdots$
represents terms of higher order in $1/L$, including $1/L\ln L$ terms
as well, and $v_{\RM c}$ is the charge velocity.
Selecting a point $(\JK,\JH)=(3t,1.65t)$
at which the spin
gap is reasonably large
we determined $E_L(Q)$ by the DMRG method for
wide ranges of $Q$ for $L=32$ and 64,
and find that the finite-size scaling is
well verified; both chain lengths studied give
$v_{\RM c}/(\vF^0 4\pi R^2)=0.082\pm3\%$ where
$\vF^0=v_{\cond}=2t\sin(\pi n/2)$ is the charge, or Fermi, velocity
in the non-interacting case.

In conclusion, we have verified that the SG phase of the
HKLM exists over an extended region of parameter space interpolating
between the large $\JK$ region where it follows from previous work on
the \tJ\ model to the weak $\JK$ region where it is predicted by
bosonization.  We have verified the expected exponential dependence of
the gap on $v/\JK$ in the weak coupling limit.  So far, we have found
no evidence for a spin gap in the pure KLM but further numerical work
is required (and is in progress) to settle this question one way or
the other.

We would like to thank J. Gan, C.S. Hellberg, A.J. Millis, and M.
Oshikawa for helpful discussions, and S. Moukouri for sharing some
numerical results. This research was supported in part by NSERC of
Canada (A.E.S. and I.A.).  A.E.S. acknowledges support from the Izaak
Walton Killam Memorial Foundation.  S.R.W. acknowledges support from
the NSF under Grant No.\ DMR-9509945.


\end{multicols}
\end{document}